\documentclass[aps,twocolumn,groupedaddress,prb,floatfix,showpacs]{revtex4}
\usepackage{mathrsfs}
\usepackage{amsmath}
\usepackage[dvips]{graphics,color}
\usepackage{graphicx}
\usepackage{epstopdf}
\usepackage{dcolumn}
\usepackage{bm}
\usepackage{longtable}
\usepackage{graphics}
\usepackage{amssymb}
\usepackage{xspace}
\usepackage{epsfig}
\usepackage{dcolumn}
\usepackage[FIGTOPCAP,hang,raggedright,nooneline,bf]{subfigure}

\begin{document}
\title{Band Structure of ABC-Stacked Graphene Trilayers}
\author{Fan Zhang$^1$}
\email{zhangfan@physics.utexas.edu}
\author{Bhagawan Sahu$^2$}
\author{Hongki Min$^{3,4}$}
\author{A.H. MacDonald$^1$}
\affiliation{
$^1$ Department of Physics, University of Texas at Austin, Austin TX 78712\\
$^2$ Microelectronics Research Center, University of Texas at Austin, Austin TX 78758\\
$^3$ Maryland NanoCenter, University of Maryland, College Park, MD 20742 \\
$^4$ Center for Nanoscale Science and Technology, National Institute of Standards and Technology, Gaithersburg, MD 20899-6202}

\date{\today}
\begin{abstract}
The ABC-stacked $N$-layer-graphene family of two-dimensional electron systems is
described at low energies by two remarkably flat bands with Bloch states that
have strongly momentum-dependent phase differences between carbon $\pi$-orbital amplitudes on
different layers, and large associated momentum space Berry phases.  These properties are most easily understood using a simplified model
with only nearest-neighbor inter-layer hopping which leads to gapless semiconductor electronic structure,
with $p^N$ dispersion in both conduction and valence bands.  We report on a study of
the electronic band structures of trilayers
which uses {\it ab initio} density functional theory and $\bm{k} \cdot \bm{p}$ theory to fit
the parameters of a $\pi$-band tight-binding model.  We find that when remote interlayer hopping is retained,
the triple Dirac point of the simplified model is split into three single Dirac points
located along the three $KM$ directions.  External potential differences between top and bottom layers
are strongly screened by charge transfer within the trilayer, but still open an energy gap at
overall neutrality.
\end{abstract}
\pacs{73.43.Cd, 71.15.-m, 71.20.-b, 81.05.ue} \maketitle

\section{Introduction}
Success\cite{first} in isolating nearly perfect monolayer and few layer sheets from bulk graphite,
along with progress\cite{First_MRSBull} in the epitaxial growth of few-layer graphene samples,
has led to an explosion of experimental and theoretical\cite{graphene_reviews_1,graphene_reviews_2,graphene_reviews_3,graphene_reviews_4,graphene_reviews_5}
interest in this interesting class of quasi-two-dimensional electron systems (2DES's).
Unique aspects of the electronic structure of graphene based 2DES's have raised
a number of new fundamental physics issues and raised hope for applications.

Monolayer graphene has a honeycomb lattice structure and is a gapless semiconductor.  Hopping between its equivalent $A$ and $B$ sublattices
gives rise to a massless Dirac fermion band structure with $J=1$ chirality when the sublattice degree of freedom is treated as a pseudospin. In
this paper we will find it useful to view the quantum two-level degree of freedom associated with two sublattice sites as a pseudospin in the
multi-layer case as well.  In AB-stacked graphene bilayers, for example, electrons on the $A_2$ and $B_1$ sublattices are repelled from the
Fermi level by a direct interlayer tunneling process with energy $\gamma_1$, leaving\cite{McCann_2006_prl} only states that are concentrated on
the $A_1$ and $B_2$ sites in the low-energy band-structure projection.  When direct hopping between $A_1$ and $B_2$ sites is neglected, the
two-step hopping process via high energy sites leads to $p^2$ conduction and valence band dispersions and to a pseudospin chirality that is
doubled,  {\em i.e.} to a phase difference between sublattice projections which is proportional to $2 \phi_{\bm{p}}$ where $\phi_{\bm{p}}$ is
the two-dimensional momentum orientation.  Pseudospin chirality has a substantial influence on interaction physics\cite{chirality_correlations}
in both single-layer and bilayer graphene, and through the associated momentum space Berry phases also on Landau quantization and the integer
quantum Hall effect.\cite{mono_pi_1,mono_pi_2,bilayer_hall, McCann_2006_prl,chiral_decomp}.  Because the two low-energy sublattices in bilayer
graphene are located on opposite layers it is possible to introduce\cite{SPBilayer_1,min_prb_2007,lu_2006,Aoki_2007,Gava_2009} a
gap\cite{spectroscopic_Ohta,spectroscopic_Kim,spectroscopic_Martin,spectroscopic_Geim,spectroscopic_Wang,spectroscopic_Heinz,transport_1,transport_2}
in the electronic structure simply by using gates to induce a difference in electric potential between layers. According to some theories a
small gap could even emerge spontaneously\cite{Zhang_2010,Guinea_physics,mft,Levitov,Vafek,Fradkin} in neutral graphene bilayers with weak
disorder because of layer inversion symmetry-breaking.

Graphene bilayer 2DES's are quite distinct from single layer 2DES's because of their flatter band dispersion and the possibility of using
external potentials to create gaps. Among all stacking possibilities, only the ABC arrangement (see below) maintains the following features that
make Bernal bilayer electronic structure interesting in thicker $N$-layer films: (i) there are two low-energy sublattice sites, implying that a
two band model provides a useful tool to describe its physics; (ii) the low energy sublattice sites are localized in the outermost layers, at
$A_1$ and $B_N$, and can be separated energetically by an electric field perpendicular to the film; (iii) hopping between low energy sites via
high energy states is an $N$-step process which leads to $p^N$ dispersion in conduction and valence bands, sublattice pseudospin chirality $N$
and Berry phase $N\pi$. The low-energy bands are increasingly flat for larger $N$, at least when weak remote hopping processes are neglected,
and the opportunity for interesting interaction and disorder physics is therefore stronger. Consequently, in the simplified chiral model, the
density-of-states $D(E)\sim E^{(2-N)/N}$ diverges as $E$ approaches zero for $N>2$ whereas it remains finite for $N=2$ and vanishes for $N=1$.
These properties also have some relevance to more general stacking arrangements since the low energy Hamiltonian of a multilayer with any type
of stacking can always be chiral-decomposed to a direct sum of ABC-stacked layers.\cite{chiral_decomp}

ABC-stacked multilayers are the chiral generalizations of monolayer and Bernal bilayer graphene, and we refer them collectively as the chiral 2D
electron system (C2DES) family.  We believe that they are likely to prove to be fertile ground for new physics.  As a first step in the
exploration of these materials we report in this paper on an effort to characterize the way in which the chirality $N$ bands of an $N$-layer
C2DES are altered by remote hopping processes neglected in the simplified model, focusing on the $N=3$ trilayer case. We use \textit{ab initio}
density functional theory (DFT) calculations, combined with a $\bm{k} \cdot \bm{p}$ expansion of the low-energy bands near the Dirac point, to
fit the parameters of a phenomenological tight-binding method (PTBM) for the $\pi$-bands of multilayer
graphene.\cite{chiral_decomp,ABC,Henrard,mccann_2009_ABC3,koshino_2010,min_2010}  We find that details of the low-energy band dispersion can be
used to fix rather definite values for the model's remote inter-layer hopping parameters.

Our paper is organized as follows.  In section II we first sketch the derivation of the low energy effective band Hamiltonian of a trilayer,
reserving details to an Appendix and explain how the interlayer hopping parameters influence the shape of constant energy surfaces.  The values
for these parameters obtained by fitting to our DFT calculations are surprisingly different from the values for the analogous hopping parameters
in Bernal stacked layers, and are not yet available from experiment. In Section II we also discuss the evolution of constant energy surface
pockets with energy, concentrating on the Lifshitz transitions at which pockets combine, in terms of Berry phase considerations and a competition between chiral
dispersion and trigonal warping.  In section III we use DFT to estimate the dependence of the trilayer energy gap on the external
potential difference between top and bottom layers and compare with predictions based on the simplified two-band model. The simplified model picture
is readily extended to higher $N$ and we use it to discuss trends in thicker ABC multilayers.  Finally, we conclude in Section IV with a
discussion of how Berry phases modify the integer quantum Hall effect and weak localization in C2DES's and with some speculations on the role of
electron-electron interactions in these two-dimensional materials.

\section{Effective model and band structure}
\subsection{Low Energy Effective Model}
\begin{figure}[htbp]
\subfigure[\color{white}{a}] {\scalebox{0.50} {\includegraphics*[2.65in,6.12in][5.95in,9.38in]{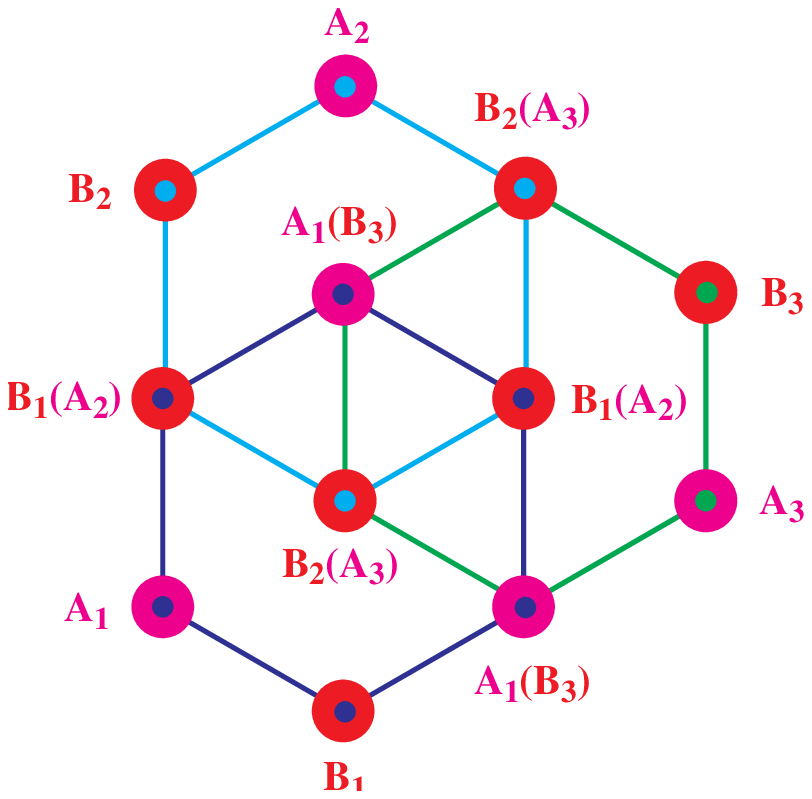}}} 
\subfigure[\color{white}{a}]{\scalebox{0.58} {\includegraphics*{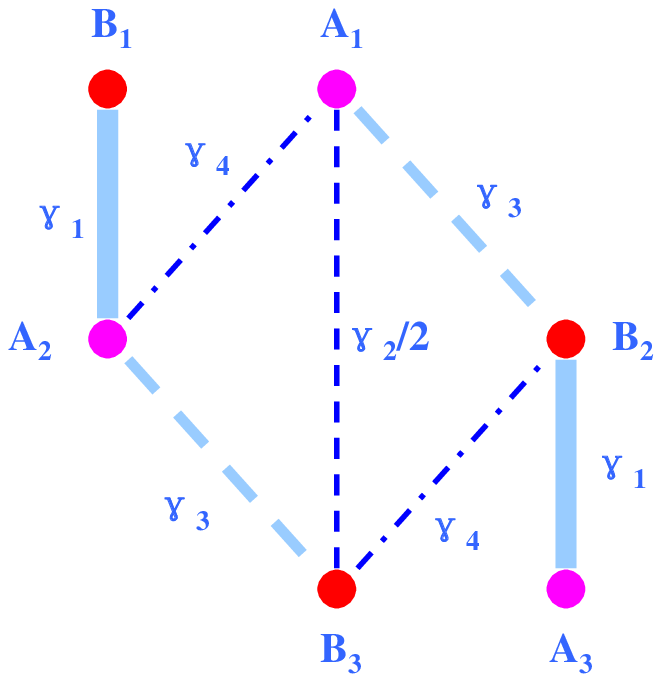}}} \caption{\label{fig:one} {(Color online) (a) Lattice structure of
ABC-stacked graphene trilayer; blue/cyan/green indicate links on the top/middle/bottom layers while purple/red distinguish the A/B sublattices.
(b) Schematic of the unit cell of ABC-stacked graphene trilayer and the most important interlayer hopping processes. }}
\end{figure}

In ABC-stacked graphene layers, each layer has inequivalent triangular $A$ and $B$ sublattices.  As illustrated in Fig.~\ref{fig:one}(a), each
adjacent layer pair forms a AB-stacked bilayer with the upper $B$ sublattice directly on top of the lower $A$ sublattice, and the upper $A$
above the center of a hexagonal plaquette of the layer below.  Our microscopic analysis uses the categorization of interlayer hopping processes
illustrated in Fig.~\ref{fig:one}(b), which is analogous to the Slonczewski-Weiss-McClure (SWM) parametrization of the tight binding model of
bulk graphite with the Bernal stacking order.\cite{parameter} Following convention $\gamma_0$ and $\gamma_1$ describe nearest neighbor
intralayer and interlayer hopping respectively, $\gamma_3$ represents hopping between the low energy sites of a AB-stacked bilayer ({\em i.e.}
$A_{\rm i} \leftrightarrow B_{\rm i+1}\;(i=1,2)$), $\gamma_4$ couples low and high energy sites located on different layers ({\em i.e.} $A_{\rm
i}\leftrightarrow A_{\rm i+1}$ and $B_{\rm i}\leftrightarrow B_{\rm i+1}\;(i=1,2)$).  We use $\gamma_2$ to denote direct hopping between the
trilayer low energy sites, and $\delta$ as the on-site energy difference of $A_1$ and $B_3$ with respect to the high energy sites.  $\gamma_5$
and $\gamma_6$ correspond to the presumably weaker couplings $B_1 \leftrightarrow A_3$ and ${\rm S}_1 \leftrightarrow {\rm S}_3\;({\rm S}=A,B)$,
respectively and $u_{\rm i}$ is used to denote the average potential of the $i$th layer.

The massless Dirac-Weyl quasiparticles of monolayer graphene are described by a ${\vec{\bm k}\cdot\vec{\bm p}}$ Hamiltonian,
\begin{eqnarray}
{\hat H}&=&v_0\left( \begin{array}{cc} 0&\pi^{\dag}\\ \pi&0\\ \end{array} \right)\,,
\end{eqnarray}
where $\pi=\xi p_{\rm x}+ip_{\rm y}$ and $\xi=+(-)$ for valley $K$($K'$).  (In the rest of the paper we focus on bands near Brillouin zone
corner $K$; the general result can be obtained by setting $p_{\rm x}$ to $\xi p_{\rm x}$.) The trilayer $\pi$-bands are the direct produce of
three sets of monolayer bands, modified by the various interlayer coupling processes identified above. In a representation of sublattice sites
in the order $A_1, B_3, B_1, A_2, B_2, A_3$, the trilayer Hamiltonian near valley $K$ can then be expressed in the form:
\begin{eqnarray}
{\mathscr{\hat H}}_{\rm trilayer}^{\rm ABC}=\left( \begin{array}{cccccc}
u_1+\delta & \frac{1}{2}\gamma_2 & v_0\pi^{\dag} & v_4\pi^{\dag} & v_3\pi & v_6\pi\\
\frac{1}{2}\gamma_2 & u_3+\delta & v_6\pi^{\dag} & v_3\pi^{\dag} & v_4\pi & v_0\pi\\
v_0\pi & v_6\pi & u_1 & \gamma_1 & v_4\pi^{\dag} & v_5\pi^{\dag}\\
v_4\pi & v_3\pi & \gamma_1 & u_2 & v_0\pi^{\dag} & v_4\pi^{\dag}\\
v_3\pi^{\dag} & v_4\pi^{\dag} & v_4\pi & v_0\pi & u_2 & \gamma_1\\
v_6\pi^{\dag} & v_0\pi^{\dag} & v_5\pi & v_4\pi & \gamma_1 & u_3\\
\end{array} \right)\,,\label{eq:fullH}
\end{eqnarray}
where $v_{\rm i}=\sqrt{3} a\gamma_{\rm i}/2\hbar$ and $a=0.246${nm}.

The identification of $A_1$ and $B_3$ as the low-energy sublattice sites is made by neglecting the weaker remote interlayer hopping processes and
setting $\pi \to 0$.  We treat coupling between the low and high-energy subspaces perturbatively by writing the trilayer Greens function as
\begin{eqnarray}
\mathscr{G}=({\mathscr{\hat H}}_{\rm trilayer}^{\rm ABC}-\epsilon)^{-1}=
\left( \begin{array}{cc} H_{11}-\epsilon & H_{12}\\ H_{21} & H_{22}-\epsilon \\
\end{array} \right)^{-1}\,
\end{eqnarray}
where the indices $1$ and $2$ denote the $2 \times 2$ low-energy block and the $4 \times 4$ high-energy
block respectively. We then solve the ${\rm Schr\ddot{o}dinger}$
equation, $(\mathscr{G})_{11}^{-1}\psi_{\rm low}=0$, by using the block matrix inversion rule
$(A^{-1})_{11}=(A_{11}-A_{12}(A_{22})^{-1}A_{21})^{-1}$ to obtain
\begin{eqnarray}
\label{hlow}
\big({(H_{11}-\epsilon)-H_{12}(H_{22}-\epsilon)^{-1}H_{21}}\big)\psi_{\rm low\,(A_1,B_3)}=0\,.
\end{eqnarray}
Since we are interested in the low-energy part of the spectrum we can view $\epsilon$ as small compared to $H_{22}$.  Expanding Eq.~(\ref{hlow})
to first order in $\epsilon$, we find that $(H_{\rm eff}-\epsilon)\psi_{\rm low}=0$, where
\begin{eqnarray}
H_{\rm eff}=\big(1+H_{12}(H_{22})^{-2}H_{21}\big)^{-1}\big(H_{11}-H_{12}(H_{22})^{-1}H_{21}\big)\,.\label{eq:phys}
\end{eqnarray}
The terms in the second
parenthesis capture the leading hopping processes between low energy sites, including virtual hopping via
high-energy states, while the first parenthesis captures an energy scale renormalization by a factor
of order $1-(v_0p/\gamma_1)^2$ due to higher-order processes which we drop except in the terms which arise
from an external potential.

Using Eq.~(\ref{eq:phys}) we find that for ABC trilayer graphene
\begin{eqnarray}
{\hat H}_{\rm eff}&=&{\hat H}_{\rm ch}+{\hat H}_{\rm s}+{\hat H}_{\rm tr}+{\hat H}_{\rm gap}+{\hat H}_{\rm s}'\,,\nonumber\\
{\hat H}_{\rm ch}&=&\frac{v_0^3}{\gamma_1^2}\left( \begin{array}{cc} 0&(\pi^{\dag})^3\\\pi^3&0\\ \end{array} \right)\nonumber\\
                 &=&\frac{(v_0p)^3}{\gamma_1^2}(\cos(3\varphi_{\rm {\bf p}})\sigma_{\rm x}+\sin(3\varphi_{\rm {\bf p}})\sigma_{\rm y})\,,\nonumber\\
{\hat H}_{\rm s}&=&\bigg(\delta-\frac{2v_0v_4p^2}{\gamma_1}\bigg)\sigma_{\rm 0}\,,\nonumber\\
{\hat H}_{\rm tr}&=&\bigg(\frac{\gamma_2}{2}-\frac{2v_0v_3p^2}{\gamma_1}\bigg)\sigma_{\rm x}\,,\nonumber\\
{\hat H}_{\rm gap}&=&u_{\rm d}\bigg(1-\bigg(\frac{v_0p}{\gamma_1}\bigg)^2\bigg)\sigma_{\rm z}\,,\nonumber\\
{\hat H}_{\rm s}'&=&\frac{u_{\rm a}}{3}\bigg(1-3\bigg(\frac{v_0p}{\gamma_1}\bigg)^2\bigg)\sigma_{\rm 0}\,. \label{eq:triH}
\end{eqnarray}
Here we have chosen $\tan\varphi_{\rm {\bf p}}=p_{\rm y}/p_{\rm x}$, defined $u_{\rm d}=(u_1-u_3)/2$ and $u_{\rm a}=(u_1+u_3)/2-u_2$, and
neglected an overall energy scale associated with the external potentials.  $\sigma_0$ is the identity matrix and the $\sigma_i$'s are Pauli
matrices acting on the low-energy pseudospin.  We have retained leading terms with cubic, quadratic, and constant dispersions, which are due
respectively to three-step, two-step, and one-step hopping processes between low energy sites.  For trilayer graphene, the linear term is absent
because the one step hopping ($\gamma_2$) is normal to the 2D space and therefore independent of momentum. ${\hat H}_{\rm ch}$ is the only term
which appears in the effective Hamiltonian in the simplified model with only nearest neighbor inter-layer tunneling.  This term has pseudospin
chirality $J=3$ and dominates at larger values of $p$. It reflects coupling between low energy sites via a sequence of three nearest neighbor
intralayer and interlayer hopping events. ${\hat H}_{\rm tr}$ is proportional to $\sigma_x$ and, because it is isotropic in 2D momentum space,
is responsible for trigonal warping of constant energy surfaces when combined with the $J=3$ chiral term. Notice that the direct hopping
$\gamma_2$ process opens a small gap at the $K$ points so that ${\hat H}_{\rm tr}$ vanishes at finite $p$ if $\gamma_2$ is positive.  ${\hat
H}_{\rm s}$ arises from a weaker coupling between low energy and high energy states that is present in bilayers and for any $N>1$ multilayer
system. This term in the effective Hamiltonian preserves layer inversion symmetry.  ${\hat H}_{\rm gap}$ captures the external potential
processes which break layer inversion symmetry and introduce a gap between electron and hole bands.  The possibility of opening a gap with an
external potential is unique to ABC stacked multilayers, increasing the possibility that they could be useful materials for future semiconductor
devices.  The strength of the gap term decreases with increasing momentum (since $v_0p\ll\gamma_1$) so that the gap around $K$ has a Mexican hat
shape, as we will discuss later. ${\hat H}_{\rm s}'$ is non-zero when the potential of the middle layer deviates from the average of the
potentials on the outermost layers. Unlike ${\hat H}_{\rm gap}$, this term preserves the layer inversion symmetry and is not responsible for an
energy gap.  A non-zero ${\hat H}_{\rm s}'$ is relevant when the electric fields in the two inter-layer regions are different. Further
discussion on the derivation of this effective Hamiltonian and on the physical meaning of the various terms can be found in the Appendix.  Note
that for strict consistency the constant terms $\delta$ and $\gamma_2/2$ should be accompanied by the factor $1-(v_0p)^2/\gamma_1^2$ based on
Eq.~(\ref{eq:phys}) which does appear in ${\hat H}_{\rm gap}$. However we ignore this factor because $\delta$ and $\gamma_2/2$ are already
small.

\subsection{{\it AB Initio} Density Functional Theory Calculations}
\begin{figure}[htbp]{\scalebox{0.37}
{\includegraphics*{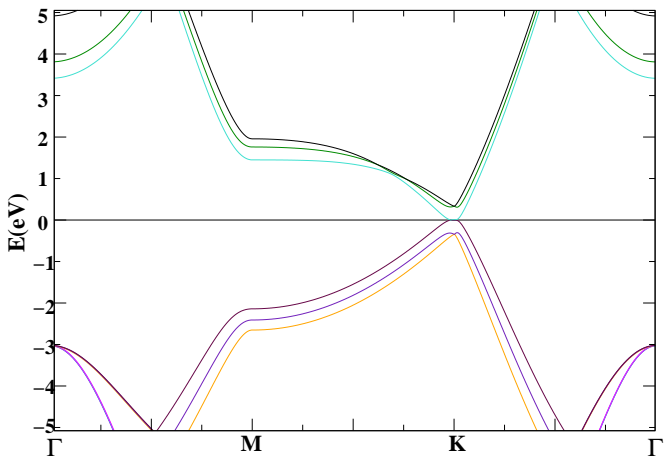}}}\caption{\label{fig:two}{(Color online) Band structure of ABC-stacked graphene trilayers in the absence of an
external electric field. The zero of energy in this plot is at the Fermi energy of a neutral trilayer.  Notice the single low-energy
band with extremely flat dispersion near the $K$ point.}}
\end{figure}
We have performed \textit{ab initio} DFT calculations\cite{espresso} for an isolated graphene trilayer in the absence of a transverse external
electric field which induces an electric potential difference between the layers. (DFT calculations in the presence of electric fields will be
discussed in the next section.) Our electronic structure calculations were performed with plane wave basis sets and ultrasoft pseudopotentials
\cite{ultrasoft}. The local density approximation (LDA) was used for the exchange and correlation potential. We fixed the interlayer separation
to 0.335 nm and placed bulk trilayer graphene in a supercell with a 40 nm vacuum region, large enough to avoid intercell interactions. A 21
$\times$ 21 $\times$ 1 {\bf k}-point mesh in the full supercell Brillouin zone (FBZ) was used with a 408 eV kinetic energy cut-off. The
calculations were tested for large {\bf k}-point meshes in the FBZ and large energy cut-offs for convergence studies. Fig.~\ref{fig:two} shows
the DFT energy band structure of ABC stacked trilayer graphene in the absence of an external electric field. The low energy band dispersion is
nearly cubic at the two inequivalent corners $K$ and $K'$ of the hexagonal Brillouin zone, as predicted by the $\pi$-orbital tight-binding and
continuum model phenomenologies. The conduction and valence bands meet at the Fermi level. Close enough to Fermi level the band is nearly flat,
which indicates the important role interactions might play in this material.

\subsection{Extracting hopping parameters from DFT}
Previously, bulk graphite (with the Bernal stacking order) SWM hopping parameters have been extensively studied using DFT and measured in
experiments. However, the values of the SWM parameters appropriate for ABC-stacked trilayer graphene were previously unknown.  We extract their
values by fitting the effective model with the DFT data in the zero electric field limit. The eigenenergies of the Hamiltonian in
Eq.~(\ref{eq:triH}) in the absence of external potentials are
\begin{equation}
\label{epmEzero}
E^{(\pm)}=h_{\rm s}\pm\sqrt{h_{\rm ch}^2+h_{\rm tr}^2+2\cos(3\varphi_{\rm{\bf p}})h_{\rm ch}h_{\rm tr}}\,,
\end{equation}
where $h_{\rm ch}=(v_0p)^3/\gamma_1^2$, $h_{\rm tr}=\gamma_2/2-2v_0v_3p^2/\gamma_1$ and $h_{\rm s}=\delta-2v_0v_4p^2/\gamma_1$.
\begin{figure}[htbp]{\scalebox{0.45}{\includegraphics*{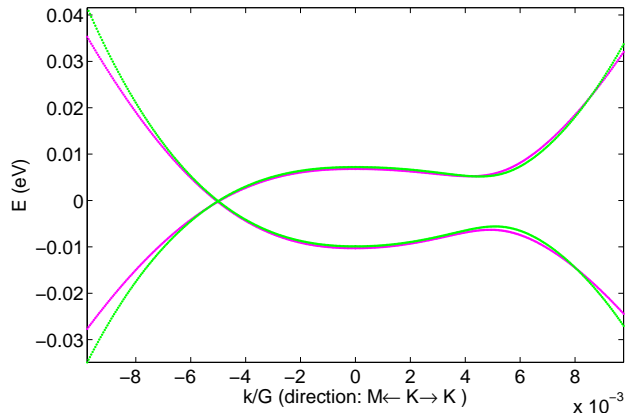}}}\caption{\label{fig:fit} {(Color online) The magenta curve is the DFT data while the Green one represents the
effective model using the extracted parameters shown in Table~\ref{table:one}. $G=4\pi/(\sqrt{3}a)$ is the length of the reciprocal vectors and
$k=0$ is the K point.}}
\end{figure}
To extract the remote hopping parameters we first set the nearest neighbor in-plane hopping parameter $\gamma_0$ to $3.16\,eV$ to set the
overall energy scale. The values of $\delta$ and, up to a sign, $\gamma_2$ can then be obtained by comparing the band energies at $p=0$
calculated by the two different methods.  Then comparing $E^{(+)}+E^{(-)}$ from the DFT data with Eq.~(\ref{epmEzero}), we obtain a value for
$\gamma_4\gamma_0/\gamma_1$.  Finally we notice that Eq.~(\ref{epmEzero}) implies that the gap between conduction ($+$) and valence ($-$) bands
vanishes at $\cos(3\varphi_{\rm{\bf p}})=1$ if $h_{\rm tr}$ is negative and at $\cos(3\varphi_{\rm{\bf p}})=-1$ if $h_{\rm tr}$ is positive.
Because of this property the Fermi level of a neutral balanced ABC trilayer is at the energy of three distinct Dirac points which are removed
from the Dirac point separated in direction by $2\pi/3$.  The triple Dirac point of the trilayer's simplified model is split into three separate
single Dirac points.  The DFT theory result that the conduction valence gap vanishes along the $K'M$ directions for which
$\cos(3\varphi_{\rm{\bf p}})=1$ implies that $h_{\rm tr}$ is negative and helps to fix the sign of $\gamma_2$.   Values for
$\gamma_3\gamma_0/\gamma_1$ and $\gamma_0^3/\gamma_1^2$ are provided by the value of $p$ at the Dirac points and the size of the splitting
between conduction and valence bands ($2\sqrt{h^2_{\rm ch}+h^2_{\rm tr}}$) along the $\cos(3\varphi_{\rm{\bf p}})=0$ directions. The best
overall fit we obtained to the bands around the $K$ point and the deformed Dirac cones is summarized in Table~\ref{table:one}, where we compare
with the corresponding fitting parameters for bulk graphite.\cite{graphene_reviews_4,parameter} Our fit is extremely good in the low energy
region in which we are interested, as shown in Fig.~\ref{fig:fit}, though there are still discrepancies as higher energies are approached. These
discrepancies are expected because of the perturbative nature of the effective model and can be partly corrected by restoring the
$1-(v_0p)^2/\gamma_1^2$ correction factor in Eq.~(\ref{eq:phys}).
\begin{table}[h]
\caption{Summary of SWM hopping parameters obtained by fitting DFT bands in ABC-stacked trilayer graphene to a low-energy effective model.  We
compare with bulk graphite values from References.\cite{graphene_reviews_4,parameter} } \centering
\newcommand\T{\rule{0pt}{3.1ex}}
\newcommand\B{\rule[-1.7ex]{0pt}{0pt}}
\begin{tabular}{c | c | c }
\hline\hline parameters & graphite($eV$)\T & ABC trilayer ($eV$)\\[3pt]
\hline $\delta$  & $0.008$\T & $-0.0014$ \\[3pt]
\hline $\gamma_1$ & $0.39$\T & $0.502$ \\[3pt]
\hline $\gamma_3$ & $0.315$\T & $-0.377$ \\[3pt]
\hline $\gamma_4$ & $-0.044$\T & $-0.099$ \\[3pt]
\hline $\gamma_2$ & $-0.020$\T & $-0.0171$ \\[3pt]
\hline\hline
\end{tabular}
\label{table:one}
\end{table}

\subsection{Electron(Hole) Pockets and Lifshitz Transitions}
\begin{figure}[htbp]
\subfigure[\color{white}{a}][{} Conduction band constant energy surfaces of an ABC graphene trilayer.]{\scalebox{0.43} {\includegraphics*[0.3in,1.8in][8.2in,9.0in]{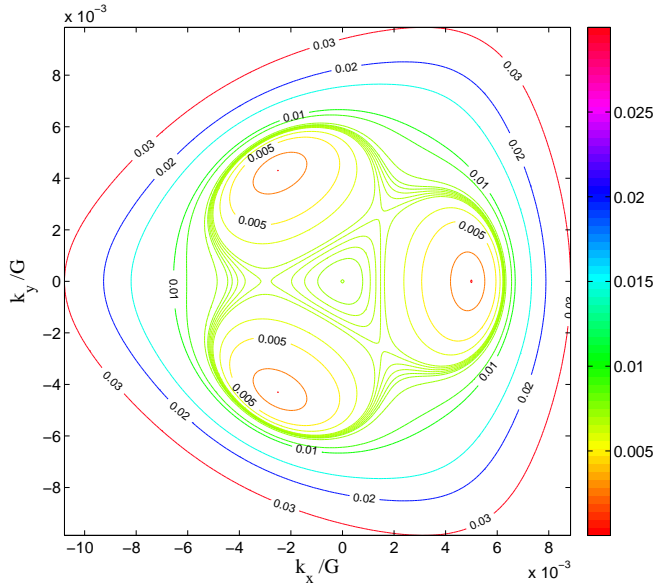}}} 
\subfigure[\color{white}{a}][{} Conduction band constant energy surfaces of an ABC graphene trilayer model with bulk graphite
parameters.]{\scalebox{0.43} {\includegraphics*[0.3in,1.6in][8.2in,9.3in]{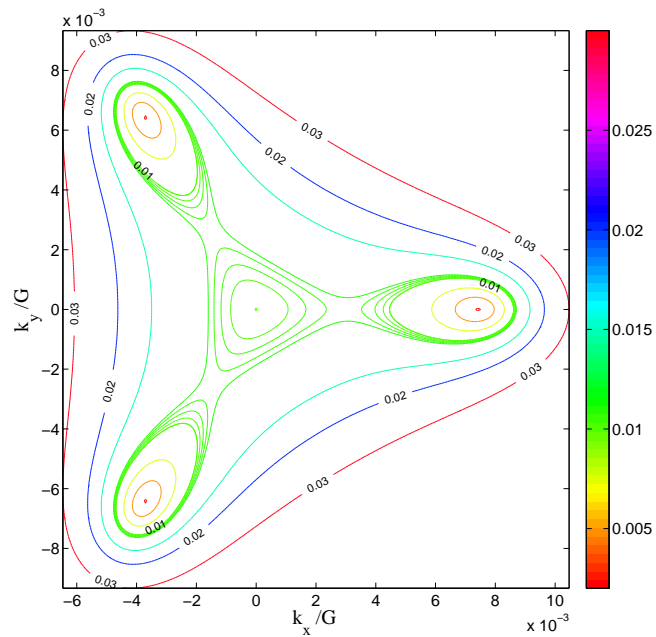}}} \caption{\label{fig:contourplot} {(Color online) Constant
energy (in units of eV) contour plots of the conduction band near zero energy. $G=4\pi/(\sqrt{3}a)$ is the length of the reciprocal vector and
$k=0$ is a K point. (a)ABC-valued, (b)Bulk graphite valued Fermi surfaces of a ABC trilayer. (a)The energies of the initial three electron
pockets from inner to outer are 0.0, 2.5, 5.0, 6.0, and 6.7 meV; The energies of the central triangles from outer to inner are 6.8, 6.9, 7.0,
7.1 and 7.2 meV; The energies of the bigger triangles from inner to outer are 6.8, 6.9, 7.0, 7.1, 7.2, 7.5, 9.0, 10.0, 15.0, 20.0, and 30.0 meV.
(b)The energies of the initial three electron pockets from inner to outer are 1.0, 5.0, 7.5, 10.0, 10.2, 10.4 and 10.6 meV; The energies of the
central triangles from inner to outer are 10.0, 10.2, 10.4 and 10.6 meV; The energies of the bigger triangles from inner to outer are 10.8,
15.0, 20.0, and 30.0 meV.}}
\end{figure}
\begin{figure}[htbp]{\scalebox{0.5}
{\includegraphics*{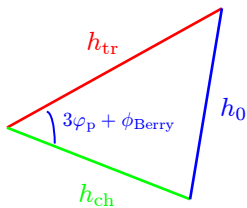}}}\caption{\label{fig:five} {(Color online) A momentum-dependent trigonometric relationship which
describes how the shape of the constant energy surfaces near the Lifshitz transitions is
collectively governed by chiral dispersion, trigonal warping, and Berry phases.}}
\end{figure}

With the effective model hopping parameters extracted from DFT we study the shape of the Fermi surface of a graphene trilayer.
Fig.~\ref{fig:contourplot}(a) shows the constant energy contour plot of the electron band around zero energy. Clearly, under remote hopping the
J=3 Dirac points evolve into three separate $J=1$ Dirac points symmetrically shifted away a little bit in the $KM$ directions ($\hat{k_{\rm
x}}$); each shifted Dirac point resembles a linear cone like the ones in monolayer graphene.  The property that total chirality is conserved can
be established by evaluating Berry phases along circular paths far from the Dirac points where the remote hopping processes do not play an
essential role.  The Dirac point distortion occurs because the direct hopping $\gamma_2$ process does not involve 2D translations and therefore
gives a momentum independent contribution to the Hamiltonian which does not vanish at the Brillouin-zone corners.  A similar distortion of the
simplified-model ideal chirality Dirac point occurs in any $3m$-layer system of ABC stacked ($m$ is a positive integer) graphene sheets. Around
each deformed Dirac cone there is a electron (hole)-like pocket in the conduction(valence) band at low carrier densities and two Lifshitz
transitions\cite{lifshitz} as a function of carrier density. Take the conduction band for example. As shown in Fig.~\ref{fig:contourplot}(a),
immediately above zero energy, the constant energy surface consists of three separate Dirac pockets. At the first critical energy $6.7$ meV, the
three electron pockets combine and a central triangle-like hole pocket appears. (Energies are measured from the Fermi energy of a neutral
trilayer.) At this energy three band-structure saddle points occur midway between the shifted Dirac points, and thus the density-of-states
diverges. Fermi levels close to these 2D logarithmic van Hove singularities could lead to broken symmetry states. At the second critical energy
$7.2$ meV, the central pocket and the three remote pockets merge into a single pocket with a smoothed triangle shape.
Fig.~\ref{fig:contourplot}(a) is in excellent agreement with constant energy surfaces constructed directly from our DFT calculations (Figure not
shown).  The two similar Lifshitz transition energies in the valence band occur at $-7.9$ meV and $-9.9$ meV. The constant energy surface at the
second Lifshitz transition solves
\begin{equation}
E^{(\pm)}({\bf p\neq 0})=E^{(\pm)}({\bf p=0})\,,
\end{equation}
where $+(-)$ refers to conduction and valence band cases. This critical condition can be specified using the law of cosines as shown in
Fig.~\ref{fig:five}, where for trilayers $\phi_{\rm Berry}=3\pi$ and $h_{0}=|\gamma_2/2|\pm 2v_0v_4p^2/\gamma_1$. This momentum-dependent
trigonometric condition can be easily generalized to the case of any other graphene multilayer and to the case with an external potential
difference.
Above the second Lifshitz transition, the constant energy surface is triangular in shape, with a trigonal distortion that differs in orientation
compared to the one obtained by plugging the bulk graphite values for the hopping parameters into the same effective model Eq.~(\ref{eq:triH})
as illustrated in Fig.~\ref{fig:contourplot}(b).  The ABC-stacked trilayer trigonal distortion has a different orientation and is weaker.  The
difference mainly reflects a difference in the sign of $\gamma_3$, which favors anti-bonding orbitals at low energies. The warping of the
constant energy surface becomes hexagonal at $8\sim 9$ meV, which provides nearly parallel flat pieces on the edges of the hexagon leading to
strong nesting. This might support some competing ground states and a density-wave ordered phase might then exist at a small but finite
interaction strength. The electronic properties of low-carrier density systems in graphene trilayers will be sensitive to these detailed band
features. Future ARPES experiments should be able to determine whether or not these features are predicted correctly by our DFT calculations.

\section{Induced Band Gaps in Trilayers}
\subsection{Energy Bands with Electric Fields}
\begin{figure}[htbp]{\scalebox{0.35} {\includegraphics*{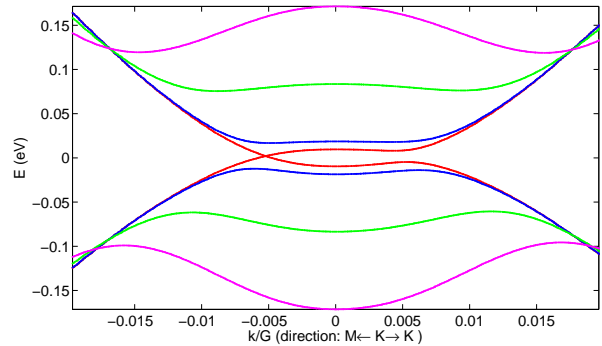}}}\caption{\label{fig:DFTE}
{(Color online) The band structures of a ABC graphene trilayer with external electric potential differences between the outermost layers.
The external potential difference $U_{\rm ext}$ values
are 0.0(red), 0.2(blue), 1.0(green) and 2.0(magenta)\,eV, respectively. $G=4\pi/(\sqrt{3}a)$ is the length of the reciprocal
vectors and $k=0$ is a K point.}}
\end{figure}

\begin{figure}[htbp]
\subfigure[\color{white}{a}][Potential Screening]{\scalebox{0.38} {\includegraphics*{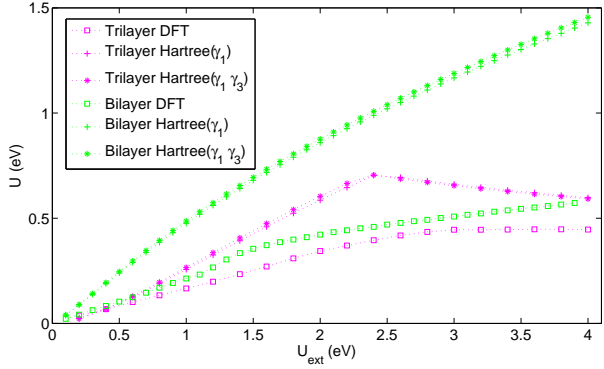}}} 
\subfigure[\color{white}{a}][Energy gap evolution]{\scalebox{0.38} {\includegraphics*{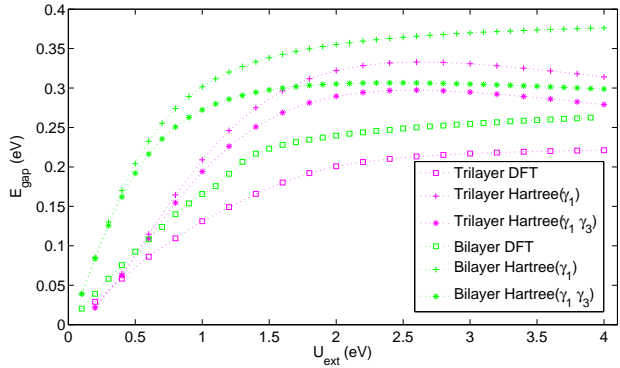}}} \caption{\label{fig:field}{(Color online)
Evolution of (a) the screened electric potential difference and (b) the energy gap, with respect to the increase of the external electric
potential difference between the outermost layers. $\Box$ represents the DFT calculations while $\dag$ ($*$) denotes the full band
self-consistent Hartree calculations without (with) remote hopping $\gamma_3$.}}
\end{figure}

Fig.~\ref{fig:DFTE} shows the energy band structure of a ABC-stacked graphene trilayer for several external electric potential differences
between the outermost layers. In the presence of an external field, as in the graphene bilayer case, the energy gap is direct but, because the
low-energy spectrum develops a \emph{Mexican hat} structure as the electric potential difference increases, occurs away from the $K$ or $K'$
point.  Charge transfer from the high-potential layer to the low-potential layer partially screens the external potential in both bilayer and
multilayer cases. Fig.~\ref{fig:field}(a) plots the screened potential $U$ and Fig.~\ref{fig:field}(b) the energy gap, as a function of the
external potential $U_{ext}$ for bilayers and trilayers calculated using both DFT and the full band self-consistent Hartree approximation. The
simple model Hartree calculations agree quite well with the DFT results generally.  We find that the screening is stronger in a trilayer system,
and that the maximum energy gap is slightly smaller. In both bilayer and trilayer, remote hopping suppresses the size of the energy gap but make
little difference to the screening.

\subsection{Self-Consistent Hartree Calculation}
As in the bilayer case, it is interesting to develop a theory of gap formation and external potential screening for ABC trilayers by combining
the low-energy effective model with a Poisson equation which takes Hartree interactions into account.  This simplified approach provides a basis
for discussing the dependence on layer number for general $N$. We therefore consider an isolated graphene $N$-layer with an interlayer
separation $d=0.335${nm} under an external electric field $E_{\rm ext}$ perpendicular to the layers, neglecting the finite thickness and
crystalline inhomogeneity of the graphene layers.  In an isolated system charge can only be transferred between layers so that $n=n_{\rm
t}+n_{\rm b}=0$. Defining $\delta n=n_{\rm b}=-n_{\rm t}$ and using a Poisson equation, we find that the screened electric potential difference
$U$ between the outermost layers is
\begin{eqnarray}
U=U_{\rm ext}+4\pi e^2(N-1)\,d\,\delta n\,. \label{eq:UUext}
\end{eqnarray}
In the two-band effective model, $\delta n$ is accumulated through the layer pseudospin polarization of the valence band states and is thus
given by the following integral over momentum space:
\begin{equation}
\delta n = \sum_{i\in v} 2\int_{BZ} {d^2 k \over (2\pi)^2} \big<\psi_{\rm i}({\bm k})\big|{\sigma_{\rm z} \over 2}\big|\psi_{\rm i}({\bm
k})\big> ,
\end{equation}
where the factor $2$ accounts for spin degeneracy, $|\psi_{\rm i}({\bm k})\rangle$ is a band eigenstate in the presence of $E_{\rm ext}$, band
index $i$ runs over all the filled valence band states, and $\sigma_z/2$ denotes the layer-pseudospin. Any Hamiltonian of a two-band model can be generally
written as $H=h_0({\bm p})+{\bm h}({\bm p})\cdot {\bm \sigma}$.  Defining $\tan\theta_{\rm {\bf p}}={\sqrt{h_1^2+h_2^2}/ h_3}$ and
$\tan\phi_{\rm {\bf p}}={h_2/ h_1}$ the conduction and valence band states in the sublattice representation are
\begin{eqnarray}
\left|+,{{\bm p}}\right>=\left(
\begin{array}{c}
{\cos{\theta_{\rm{\bf p}} \over 2}}\\
{\sin{\theta_{\rm {\bf p}} \over 2}} e^{i\phi_{\rm {\bf p}}}\\
\end{array}
\right),\quad \left|-,{\bm p}\right>= \left(
\begin{array}{c}
{-\sin{\theta_{\rm{\bf p}} \over 2}} \\
{\cos{\theta_{\rm {\bf p}} \over 2}} e^{i\phi_{\rm {\bf p}}}\\
\end{array}
\right)\,.
\end{eqnarray}
It follows that
\begin{eqnarray}
\label{eq:formula}
\delta n &=& 4 \int_{|p|<p_{\rm c}} {d^2 p \over (2\pi\hbar)^2} \big<-,{\bm p}\big|{\sigma_z \over 2}\big|-,{\bm p}\big> \nonumber \\
&=&-{1\over 2\pi^2\hbar^2}\int_0^{2\pi}\int_0^{p_{\rm c}}\cos\theta_{\rm {\bf p}}\,p\,dp\,d\varphi_{\rm {\bf p}} \,,
\end{eqnarray}
where $p_{\rm c}=\gamma_1/v_0$ is the high momentum cutoff of the effective model and $\varphi_{\rm {\bf p}}$ is the angle of ${\bm p}$.

Let's first discuss the simplified two-band model which has only the
chiral term. For general $N$
\begin{eqnarray}
{\hat H}^{\rm (N)}_{\rm ch}&=&\frac{v_0^{\rm N}}{(-\gamma_1)^{\rm N-1}}\left( \begin{array}{cc} 0&(\pi^{\dag})^N\\\pi^N&0\\ \end{array}
\right)\nonumber\\&=&\frac{(v_0p)^{\rm N}}{(-\gamma_1)^{\rm N-1}}(\cos(N\varphi_{\rm {\bf p}})\,\sigma_{\rm x}+\sin(N\varphi_{\rm{\bf
p}})\,\sigma_{\rm y})\,. \label{eq:Nlayer}
\end{eqnarray}
The electric potential in the two-band model is $\pm {U_{\rm ext}\over 2}\,\sigma_{\rm z}$. Inserting Eq.~(\ref{eq:Nlayer}) in
Eq.~(\ref{eq:UUext}) and Eq.~(\ref{eq:formula}), we obtain an algebraic formula for the self-consistent Hartree potential valid for general $N$:
\begin{eqnarray}
\frac{U_{\rm ext}}{\gamma_1}&=&\frac{U}{\gamma_1}+\frac{4(N-1)\,d}{a_0}\,\frac{m_2}{m_{\rm e}}\,F(N,{U})\,,\\
F(N,{U}) &=&\frac{1}{t_{\rm c}}\int_0^{t_{\rm c}}\frac{d\,t}{\sqrt{t^N+1}}\,\nonumber\\
&=&{}_{2}F_{1}({1\over N},{1\over 2},{1+N\over
N},-({2\gamma_1\over U})^{2})\,,\nonumber
\end{eqnarray}
where $a_0=0.053${nm} is the Bohr radius, $m_2$ is the effective mass of a graphene bilayer, $t_{\rm c}=(2\gamma_1/U)^{2/N}$ and ${}_2F_1$ is
Gauss' hypergeometric function. In the limit of large $N$, $F(N,U) \to 1$ and thus the Hartree equation reduces to
\begin{eqnarray}
U\simeq U_{\rm ext}-\frac{4(N-1)\,d}{a_0}\,\frac{m_2}{m_{\rm e}}\gamma_1\,
\end{eqnarray}
except at very small $U$.  For small $U$ and $N=2$, the Hartree equation reads
\begin{eqnarray}
\frac{U_{\rm ext}}{U}\simeq\frac{2d}{a_0}\,\frac{m_2}{m_{\rm e}}\ln\frac{4\gamma_1}{U}\,,
\end{eqnarray}
which is consistent with previous Hartree calculations in graphene bilayers\cite{min_prb_2007}. In the limit of small $U$ for
$N>2$, the Hartree equation has the asymptotic form
\begin{eqnarray}
\frac{U}{2\gamma_1}\simeq \bigg(\frac{U_{\rm ext}}{2\gamma_1}\bigg)^{\frac{N}{2}}C\,,
\end{eqnarray}
where the factor $C=[\frac{2(N-1)d}{a_0}\,\frac{m_2}{m_{\rm e}}\,(1-\frac{2}{2-N}+\frac{1}{2-3N})]^{-N/2}$. The larger the value of $N$, the
flatter the chiral bands, and the stronger the screening.   For $N=2$ the screening response
is linear up to a logarithmic factor, while for larger $N$, superlinear screening leads to a screened potential difference which
initially grows slowly with external potential following $U\propto U_{\rm ext}^{N/2}$.
The strongest possible screening reduction of the external potential
corresponds to the Hartree-potential due to transfer of all the states in the energy
regime $\leq 2 \gamma_1$ over which the low energy model applies to one layer.

\begin{figure}[htbp]{\scalebox{0.38}
{\includegraphics*{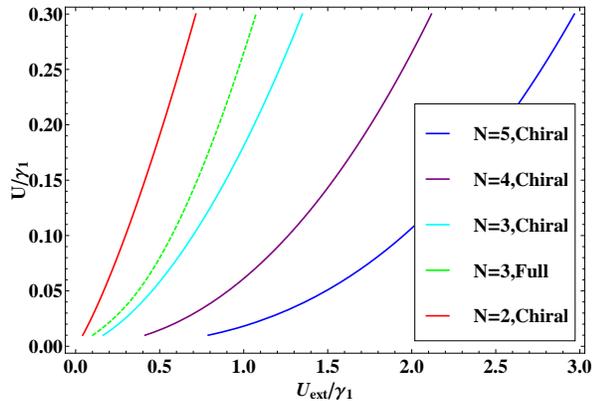}}}\caption{\label{fig:six} {(Color online) $U\;{\rm v.s.}\;U_{\rm ext}$ plot describes the screening effect in
different chiral-$N$ systems. The Chiral model results refer to the Hamiltonian in Eq.~(\ref{eq:Nlayer}) while the full model results refer to
the Hamiltonian in Eq.~(\ref{eq:triH}).}}
\end{figure}

For the trilayer case we can perform a similar calculation using the full low-energy Hamiltonian derived in Eq.~(\ref{eq:triH}).
In this case we
find that
\begin{eqnarray}
\frac{U_{\rm ext}}{\gamma_1}&=&\frac{U}{\gamma_1}+\frac{8\,d}{a_0}\,\frac{m_2}{m_{\rm e}}\,G({U})\,,\label{eq:triScreen}\\
G({U})&=&{2\over \pi}\int_0^1 dt\frac{h_{\rm gap}\, K\big(\frac{4\sqrt{2}h_{\rm ch}\,h_{\rm tr}}{(h_{\rm ch}+\sqrt{2}h_{\rm tr})^2+h_{\rm
gap}^2}\big)}{\sqrt{(h_{\rm ch}+\sqrt{2}h_{\rm tr})^2+h_{\rm gap}^2}}\,,\nonumber
\end{eqnarray}
where $h_{\rm ch}=t^{3\over 2}$, $h_{\rm tr}=|{\gamma_2\over 2\gamma_1}-{2v_3\over v_0}t|$, $h_{\rm gap}={U\over 2\gamma_1}(1-t)$, $t=({v_{\rm
0}p\over \gamma_1})^2$, and $K(x)$ is the complete elliptic integral of the first kind. Fig.~\ref{fig:six} compares the screening properties of
the full low-energy effective model for trilayers to the chiral model results for $N=2,3,4,5$.  For $U_{\rm ext}<\gamma_1/2$, the energy regime
over which the low-energy effective model applies, we see that screening increases systematically with $N$ because of smaller gaps between
conduction and valence band orbitals which make the occupied valence band orbitals more polarizable. The comparison between the simplified
chiral model and the low-energy effective model for $N=3$ demonstrates that remote hopping processes suppress screening because they tend to
increase the gap between conduction and valence bands at momenta near the Brillouin-zone corner.

In concluding this section we caution that occupied $\sigma$ orbitals, neglected in the low-energy effective model and $\pi$-band tight-binding
models, will contribute slightly to polarization by an external electric field and therefore to screening. Furthermore exchange potentials will
also be altered by an external electric field and influence the screening. Since exchange interactions are attractive, they always work against
screening and will make a negative contribution to the screening ratio we have discussed in multilayers. Because the low energy eigenstates in
multilayers are coherent superpositions of states localized in different layers, our DFT calculations which employ a local exchange
approximation, might also yield inaccurate results for the screening ratio. In fact simple measurements of the screening properties might
provide a valuable window on many-body physics in ABC-stacked graphene multilayers which lies outside the scope of commonly employed
approximations.

\section{Discussion}
We have derived an effective model for the low-energy conduction and valence bands of an ABC-stacked graphene multilayer.  The low-energy model
can be viewed as a momentum-dependent pseudospin Hamiltonian, with the pseudospin constructed from the low energy sites on the top and bottom
layers.  The simplified version of this model starts from a $\pi$-band tight-binding model with only nearest neighbor hopping and yields a
pseudospin magnetic field whose magnitude varies as momentum $p^{N}$ in an N-layer stack and whose direction is $N \phi_{\bf{p}}$ where
$\phi_{\bf p}$ is the momentum orientation. The likely importance of electron-electron interactions in multilayers can be judged by comparing
the characteristic band and interaction energies in a system with carrier density $n$ and Fermi wavevector $p_F \propto \sqrt{n}$. The
characteristic Coulomb interaction energy per-particle in all cases goes like $e^2 n^{1/2}$, while the band energy goes like $n^{N/2}$. For
low-carrier densities the band energy scale is always smaller. In the case of trilayer ABC graphene, the interaction energy scale is larger than
the band energy scale for carrier density $n <  10^{12}{\rm cm}^{-2}$.

Although interactions are clearly important and can potentially introduce new physics, the chiral band model is not valid at low-densities
because of the influence of remote hopping processes which we have estimated in this article by carefully fitting a low-energy effective model
to DFT bands.  The Hamiltonian in Eq.~(\ref{eq:triH}) combined with the parameters in Table I should be used to describe graphene trilayers with
low carrier densities.  In a realistic system the Fermi surface of a ABC trilayer with a low carrier density consists of three electron pockets
centered away from the K point. As the carrier density grows these pockets convert via a sequence of two closely spaced Lifshitz transitions
into a single K-centered pocket.  The carrier density at the Lifshitz transition is $\sim 10^{11}{\rm cm}^{-2}$, which translates to a Coulomb
interaction scale of $\sim 45$ meV, compared to a Fermi energy of $\sim 7$ meV.

The Berry phase associated with the momentum-dependence of the pseudospin orientation field, $\pi$ for a full rotation in single-layers and
$2\pi$ in the bilayer chiral model for example, is
known\cite{localization_Nagaosa,localization_Ando,localization_Guinea,localization_McCann,localization_Falko,localization_ex,localization_physics}
to have an important influence on quantum corrections to transport. Because of their very different Berry phases time-reversed paths are
expected to interfere destructively for $N$-odd systems while constructively for $N$-even system, leading to weak anti-localization for odd $N$
and weak localization for even $N$.  This general tendency will however be altered by trigonal and other corrections to the low-energy effective
Hamiltonian, like those we have derived for trilayers. The influence of these band features on quantum corrections to transport can be evaluated
starting from the results obtained here.

Another important consequence of Berry phases in the chiral model is the unconventional Landau level structure it
yields\cite{mono_pi_1,mono_pi_2,bilayer_hall,McCann_2006_prl,chiral_decomp}. In the chiral model for ABC trilayers there is a three-fold
degeneracy at the Dirac point, in addition to the usual spin and valley degeneracies.  This grouping of Landau level leads to the expectation
that quantum Hall studies in trilayers will reveal plateaus that jump from one at $-6e^2/h$ to one at $6e^2/h$. Electron-electron interactions
acting alone are expected to lift these degeneracies and give rise to quantum Hall ferromagnetism\cite{QHF_theory,QHF_ex1,QHF_ex2}.  These
interaction effects will act in concert with small corrections to the Landau level structures due to the remote hopping terms that have been
quantified in this paper.

Although we have discussed the case of ABC stacked trilayers, we expect qualitatively similar results for ABC stacking sequences of general
thickness $N$.  At low energies the band structure will consist of a conduction and a valence band with $p^{N}$ dispersion and a gap in the
presence of an external electric field across the film.  In the presence of a magnetic field $N$ Landau levels are pinned to the neutral system
Fermi level for each spin and valley.  At the lowest energies, within around $10${\rm meV} of the neutral system Fermi level, constant energy
surfaces will be strongly influenced by remote hopping processes which will also split the Dirac point Landau levels. The remote hopping terms
give rise to saddle-points in the band structure at which the density-of-states will diverge.  Broken symmetry electronic states are mostly
likely to occur when the Fermi level is coincident with these saddle points. The energy range over which the low-energy effective model applies
will, however, decrease with film thickness.  We expect both disorder and interaction effects to be strong within this family of low-dimensional
electron systems, which should be accessible to experimental study in samples for which disorder is weak on the energy scale over which the
low-energy effective model applies.

In summary, we have derived an effective model for trilayers, extracted the hopping parameters for ABC-stacked multilayers, from DFT and studied
the trilayer Fermi surfaces.  Furthermore, we have explored the screening effect in trilayers and then explained and compared with other C2DES
cases by a tight-binding model self-consistent Hartree method.  Lastly, we have argued the importance of Berry phases and interactions in C2DES.

\section{Acknowledgements} Work by F. Z. and A. H. M. was supported by the Welch Foundation, by the NSF under grant DMR-0606489, and by the DOE
under grant DE-FG03-02ER45985.  F. Z., B. S., and A. H. M. acknowledge financial support from the NRI-SWAN program. Work done by H. M. has been
supported in part by the NIST-CNST/UMD-NanoCenter Cooperative Agreement.  We acknowledge the allocation of computing time on NSF Teragrid
machine {\it Ranger} (TG-DMR090095) at the Texas Advanced Computing Center.  The authors thank J. J. McClelland, M. D. Stiles and E. J. Cockayne
for their valuable comments.

\appendix
\section{Diagrammatic Derivation of the Low Energy Effective Model}
\begin{figure}[htp]
\subfigure[\color{white}{a}]{\scalebox{0.45}{\includegraphics*{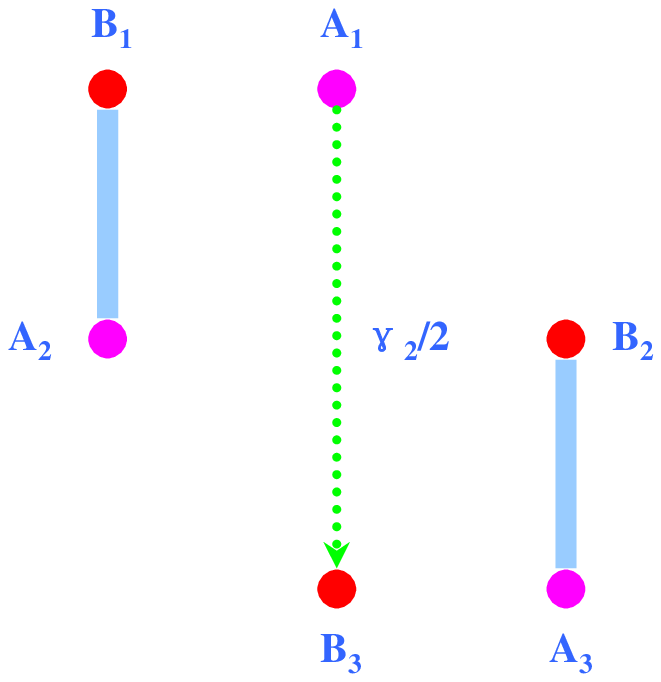}}}\hspace{0.6in} \subfigure[\color{white}{a}]
{\scalebox{0.45}{\includegraphics*{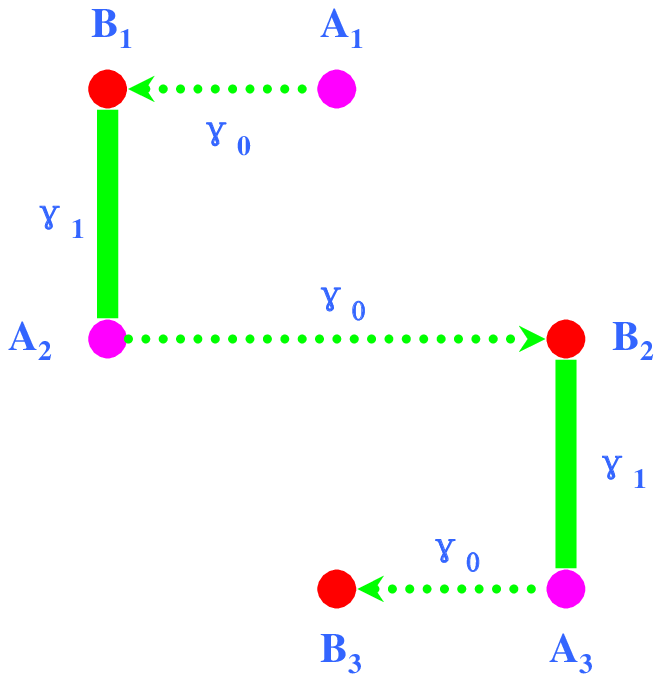}}} \subfigure[\color{white}{a}] {\scalebox{0.45} {\includegraphics*{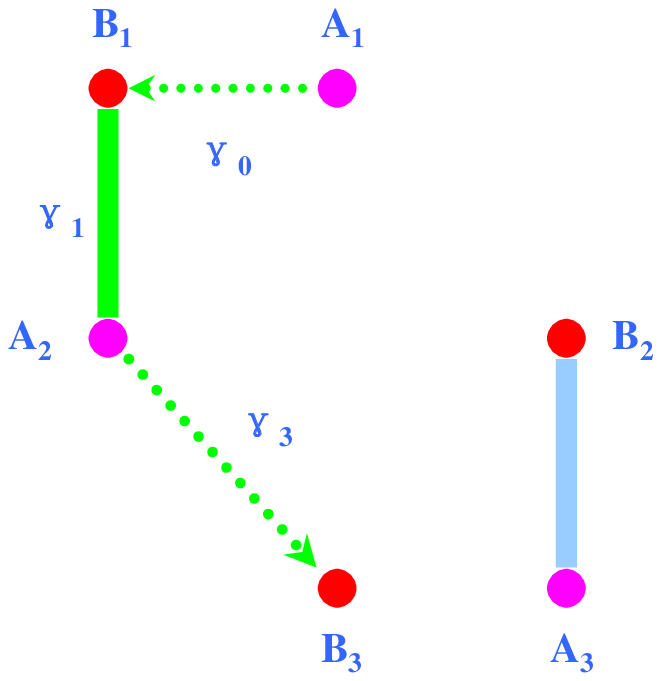}}}\hspace{0.6in}
\subfigure[\color{white}{a}] {\scalebox{0.45} {\includegraphics*{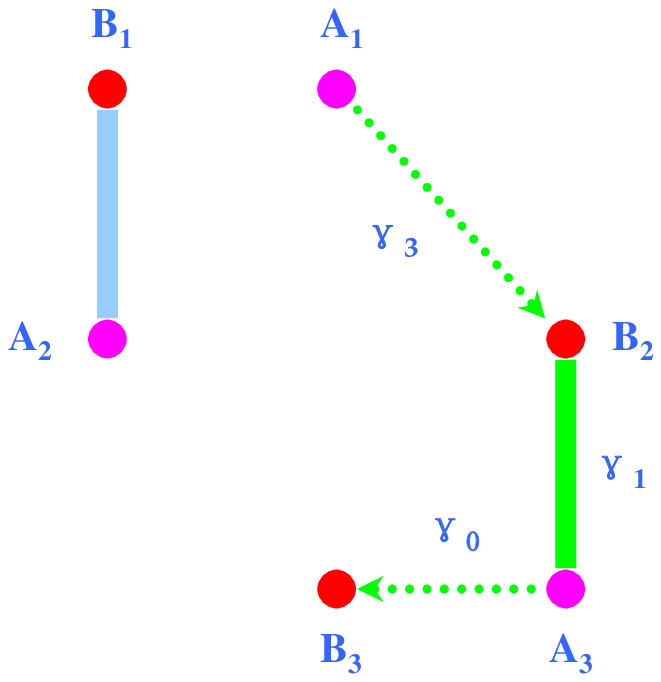}}} \subfigure[\color{white}{a}] {\scalebox{0.45}
{\includegraphics*{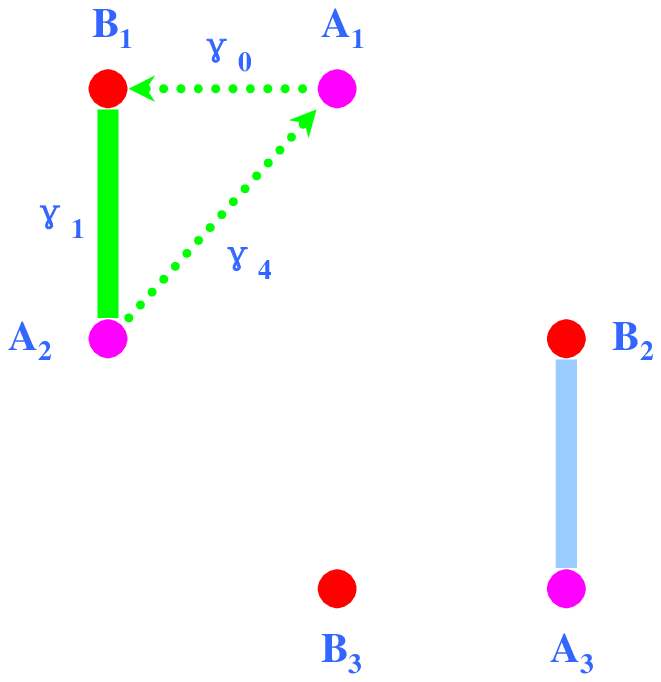}}} \hspace{0.6in} \subfigure[\color{white}{a}]{\scalebox{0.45} {\includegraphics*{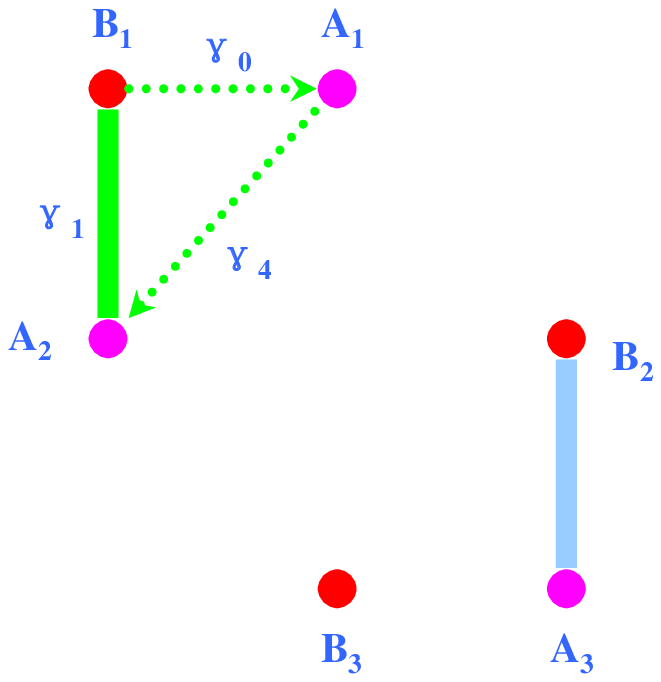}}}
\caption{\label{fig:ten} {(Color online) Schematic of hoppings from $A_1$ to $B_3$; (a) one-step $A_1\rightarrow B_3$ and (b) three-step
$A_1\rightarrow {B_1A_2}\rightarrow {B_2A_3}\rightarrow B_3$ and (c) (d) two-step $A_1\rightarrow {B_1A_2}\rightarrow B_3$ and $A_1\rightarrow
{B_2A_3}\rightarrow B_3$.  Schematic of hoppings from $A_1$ to $A_1$; (e) two-step $A_1\rightarrow {B_1A_2}\rightarrow A_1$ and (f) two-step
$A_1\rightarrow {A_2B_1}\rightarrow A_1$. }}
\end{figure}
As a result of tight-binding model, each term of the effective Hamiltonian Eq.~(\ref{eq:triH}) has a unique physical picture. Hereafter, we view
the strongly stacked pair $B_{i}A_{\rm i+1}$ as a single dimer site and assume zero external potentials for simplicity. The general formula of
effective low energy models Eq.~(\ref{eq:phys}) can be understood as following. The terms in the second parenthesis represent the leading
hopping processes, while the terms in the first parenthesis are approximately $1-(v_0p/\gamma_1)^2$ and give a small correction. $H_{11}$ is the
unperturbed Hamiltonian of low energy sites and thus includes the direct hopping and on-site energy. $H_{21}$ and $H_{12}$ are hoppings from and
to low energy sites, respectively, describing the coupling to high energy ones. $H_{22}$ contains the hoppings between high energy sites and is
an intermediate process. Therefore $H_{12}(H_{22})^{-1}H_{21}$ together gives the general ``three''-step hoppings which start from and end at
low energy sites by way of high energy ones.  Note that the intermediate process within high energy sites is zero for single layers, a constant
for bilayers, one-step for trilayers, and multi-step for $N\geq3$ layers. In bilayers for example, the linear trigonal warping term arises from
$H_{11}$, while the chiral term attributes to $H_{12}(H_{22})^{-1}H_{21}$. Because $H_{22}$ gives no hopping and is simply $\gamma_1$,
$H_{12}(H_{22})^{-1}H_{21}$ is reduced to two-step and hence the chiral term is quadratic.  In the trilayer case, for the matrix element
$B_3A_1$, $H_{11}$ provides the first term of ${\hat H}_{\rm tr}$ shown in Fig.~\ref{fig:ten}(a) while $H_{12}(H_{22})^{-1}H_{21}$ contributes
${\hat H}_{\rm ch}$ and the second term of ${\hat H}_{\rm tr}$ as depicted in Fig.~\ref{fig:ten}(b) and (c)(d), respectively.
$H_{12}(H_{22})^{-1}H_{21}$ also gives rise to the second term of ${\hat H}_{\rm s}$ for the matrix element $A_1A_1$ as presented in
Fig.~\ref{fig:ten}(e)(f).

Generally, in order to derive the low energy effective model for a general ABC-stacked $N$-layer graphene, we first need to write a $2N\times
2N$ Hamiltonian matrix as Eq.~(\ref{eq:fullH}), then we specify all the leading hopping processes in the diagrammatic language like
Fig.~\ref{fig:ten}, instead of inverting the large Hamiltonian matrix. The hopping diagrams are convenient for systematic calculations in a way
similar to the way Feynman diagrams help in perturbation theories. The exact coefficient of one hopping process can be easily calculated using
Eq.~(\ref{eq:phys}) by picking up the starting and ending sites, setting matrix elements of unrelated sites as zero and turning off the
unrelated hopping parameters. Frequently, one hopping process can be neglected because its requirement of more than one sub-hopping with
comparably small amplitudes.


\begin{thebibliography}{100}
\bibitem{first}
 K. S. Novoselov, A. K. Geim{\it et al.}, Science {\bf 306}, 666 (2004).

\bibitem{First_MRSBull} P. First {\em et al.}, MRS Bulletin, to appear April (2010).

\bibitem{graphene_reviews_1}
 T. Ando, {J. Phys. Soc. Jpn.} {\bf 74}, 777-817 (2005).

\bibitem{graphene_reviews_2}
 A. K. Geim and K. S. Novoselov, {Nature Mater.} {\bf 6}, 183 (2007).

\bibitem{graphene_reviews_3}
 A. K. Geim and A. H. MacDonald, {Phys. Today} {\bf 60}(8), 35(2007).

\bibitem{graphene_reviews_4}
 A. H. Castro Neto {\it et al.}, {Rev. Mod. Phys.} {\bf 81}, 109 (2009).

\bibitem{graphene_reviews_5}
Michael Fuhrer {\it et al.}, MRS Bulletin, to appear April (2010).

\bibitem{McCann_2006_prl}
 E. McCann and V. I. Fal'ko, {Phys. Rev. Lett.} {\bf 96}, 086805 (2006).

\bibitem{chirality_correlations}
 Y. Barlas, T. Pereg-Barnea, M. Polini, R. Asgari, A. H. MacDonald, {Phys. Rev. Lett.} {\bf 98}, 236601 (2007).

\bibitem{mono_pi_1}
 K. S. Novoselov {\it et al.}, Nature (London) {\bf 438}, 197 (2005).

\bibitem{mono_pi_2}
 Y. Zhang, Y. W. Tan, H. L. Stormer, and P. Kim, Nature (London) {\bf 438}, 201 (2005).

\bibitem{bilayer_hall}
 K. S. Novoselov {\it et al.}, Nat. Phys. {\bf 2}, 177 (2006).

\bibitem{chiral_decomp}
 H. Min and A. H. MacDonald, Phys. Rev. B {\bf 77}, 155416 (2008);
 H. Min and A. H. MacDonald, Prog. Theor. Phys. Suppl. {\bf 176}, 227 (2008).


\bibitem{SPBilayer_1}
 E. McCann, {Phys. Rev. B}, {\bf 74}, 161403(R) (2006).

\bibitem{min_prb_2007}
 H. Min, B. Sahu, S. K. Banerjee and A. H. MacDonald, {Phys. Rev. B.} {\bf 75}, 155115 (2007).

\bibitem{lu_2006}
 C. L. Lu {\it et al.}, Phys. Rev. B {\bf 73}, 144427 (2006).

\bibitem{Aoki_2007}
 M. Aoki and H. Amawashi, Solid State Commun. {\bf 142}, 123 (2007).

\bibitem{Gava_2009}
 P. Gava, M. Lazzeri, A. M. Saitta, and F. Mauri, Phys. Rev. B {\bf 79}, 165431 (2009).

\bibitem{spectroscopic_Ohta}
 T. Ohta {\it et al.}, {Science} {\bf 313}, 951 (2006).

\bibitem{spectroscopic_Kim}
 Z. Q. Li {\it et al.}, Phys. Rev. Lett. {\bf 102}, 037403 (2009).

\bibitem{spectroscopic_Martin}
 L. M. Zhang {\it et al.}, Phys. Rev. B {\bf 78}, 235408 (2008).

\bibitem{spectroscopic_Geim}
 A. B. Kuzmenko{\it et al.}, Phys. Rev. B {\bf 79}, 115441 (2009).

\bibitem{spectroscopic_Wang}
 Y. Zhang {\it et al.}, Nature (London) {\bf 459}, 820 (2009).

\bibitem{spectroscopic_Heinz}
 K. F. Mak, C. H. Lui, J. Shan, and T. F. Heinz, Phys. Rev. Lett. {\bf 102}, 256405 (2009).

\bibitem{transport_1}
 E. V. Castro {\it et al.}, Phys. Rev. Lett. {\bf 99}, 216802 (2007).

\bibitem{transport_2}
 J. B. Oostinga {\it et al.}, Nat. Mater. {\bf 7}, 151 (2007).

\bibitem{Zhang_2010}
 F. Zhang, H. Min, M. Polini and A. H. MacDonald, Phys. Rev. B {\bf 81}, 041402(R) (2010).

\bibitem{Guinea_physics}
 F. Guinea, Physics {\bf 3}, 1 (2010).

\bibitem{mft}
 H. Min, G. Borghi, M. Polini, and A. H. MacDonald, Phys. Rev. B {\bf 77}, 041407(R) (2008).

\bibitem{Levitov}
 R. Nandkishore and L. Levitov, arXiv:0907.5395 (unpublished).

\bibitem{Vafek}
 O. Vafek and K. Yang, Phys. Rev. B {\bf 81}, 041401(R) (2010).

\bibitem{Fradkin}
 K. Sun, H. Yao, E. Fradkin, and S. Kivelson, Phys. Rev. Lett. {\bf 103}, 046811 (2009).

\bibitem{ABC}
 F. Guinea, A. H. Castro Neto and N. M. R. Peres, Phys. Rev. B {\bf 73}, 245426 (2006).

\bibitem{Henrard}
 S. Latil and L. Henrard, Phys. Rev. Lett. {\bf 97}, 036803 (2006).

\bibitem{mccann_2009_ABC3}
 M. Koshino and E. McCann, Phys. Rev. B {\bf 80}, 165409 (2009).

\bibitem{koshino_2010}
 M. Koshino, arXiv:0911.3484 (unpulished) (2010).

\bibitem{min_2010}
H. Min, M. D. Stiles, and A. H. MacDonald (unpublished).

\bibitem{parameter}
 M. S. Dresselhaus and G. Dresselhaus, Adv. Phys. {\bf 51}, 1 (2002).

\bibitem{espresso}
 P. Giannozzi {\it et al.}, {\it 'Quantum ESPRESSO:  a modular and open-source software project for quantum simulation of materials'}, J. Phys: Condens. Matter {\bf 21}, 395502 (2009) and http://www.quantumespresso.org/.

\bibitem{ultrasoft}
 David Vanderbilt, Phys. Rev. B {\bf 41}, R7892 (1990).

\bibitem{lifshitz}
 L. M. Lifshitz, Zh. Eksp. Teor. Fiz. {\bf 38}, 1565 (1960).

\bibitem{localization_Nagaosa}
 S. Hikami, A. I. Larkin and N. Nagaosa, Progr. Theor. Phys. {\bf 63}, 707 (1980).

\bibitem{localization_Ando}
 H. Suzuura and T. Ando, Phys. Rev. Lett. {\bf 89}, 266603 (2002).

\bibitem{localization_Guinea}
 A. F. Morpurgo and F. Guinea, Phys. Rev. Lett. {\bf 97} 196804 (2006).

\bibitem{localization_McCann}
 E. McCann {\it et al.}, Phys. Rev. Lett. {\bf 97}, 146805 (2006).

\bibitem{localization_Falko}
 K. Kechedzhi, V. I. Fal'ko, E. McCann and B. L. Altshuler, Phys. Rev. Lett. {\bf 98}, 176806 (2007).

\bibitem{localization_ex}
 F. V. Tikhonenko, A. A. Kozikov, A. K. Savchenko, and R. V. Gorbachev, Phys. Rev. Lett. {\bf 103}, 226801 (2009).

\bibitem{localization_physics}
 E. McCann, Physics {\bf 2}, 98 (2009).

\bibitem{QHF_theory}
 Y. Barlas, R. Cote, K. Nomura and A. H. MacDonald, Phys. Rev. Lett. {\bf 101}, 097601(2008).

\bibitem{QHF_ex1}
 B. Feldman, J. Martin and A. Yacoby, Nature Phys. {\bf 5}, 889 (2009).

\bibitem{QHF_ex2}
 Y. Zhao, P. Cadden-Zimansky, Z. Jiang and P. Kim, Phys. Rev. Lett. {\bf 104}, 066801 (2010).
\end{thebibliography}
\end{document}